# Cetacean Translation Initiative: a roadmap to deciphering the communication of sperm whales

Jacob Andreas[1], Gašper Beguš[2], Michael M. Bronstein[3,4,5], Roee Diamant[6], Denley Delaney[7], Shane Gero[8,9], Shafi Goldwasser[10]*, David F. Gruber[11], Sarah de Haas[12], Peter Malkin[12], Roger Payne[13], Giovanni Petri[14], Daniela Rus[1], Pratyusha Sharma[1], Dan Tchernov[7], Pernille Tønnesen[15], Antonio Torralba[1], Daniel Vogt[16], Robert J. Wood[16,*]

[1]MIT CSAIL, Cambridge MA, USA, [2]Department of Linguistics, University of California, Berkeley, CA, USA, [3]Department of Computing, Imperial College London, United Kingdom, [4]IDSIA, University of Lugano, Switzerland, [5]Twitter, United Kingdom, [6]School of Marine Sciences, Haifa University, Israel, [7]Exploration Technology Lab, National Geographic Society, [8]Dominica Sperm Whale Project, Dominica, [9]Department of Biology, Carleton University, Canada, [10]Simons Institute for the Theory of Computing, University of California, Berkeley, CA, USA, [11]Department of Natural Sciences, Baruch College and The Graduate Center, PhD Program in Biology, City University of New York, New York, NY, USA, [12]Google Research [13]Ocean Alliance, Gloucester, MA, USA, [14]ISI Foundation, Turin, Italy, [15]Marine Bioacoustics Lab, Zoophysiology, Department of Biology, Aarhus University, Denmark, [17]School of Engineering and Applied Sciences, Harvard University, Cambridge, MA, USA

*The authors are the current scientific members of Project CETI collaboration, listed in alphabetical order.

## Abstract

The past decade has witnessed a groundbreaking rise of machine learning for human language analysis, with current methods capable of automatically accurately recovering various aspects of syntax and semantics — including sentence structure and grounded word meaning — from large data collections. Recent research showed the promise of such tools for analyzing acoustic communication in nonhuman species. We posit that machine learning will be the cornerstone of future collection, processing, and analysis of multimodal streams of data in animal communication studies, including bioacoustic, behavioral, biological, and environmental data. Cetaceans are unique non-human model species as they possess sophisticated acoustic communications, but utilize a very different encoding system that evolved in an aquatic rather than terrestrial medium. Sperm whales, in particular, with their highly-developed neuroanatomical features, cognitive abilities, social structures, and discrete click-based encoding make for an excellent starting point for advanced machine learning tools that can be applied to other animals in the future. This paper details a roadmap toward this goal based on currently existing technology and multidisciplinary scientific community effort. We outline the key elements required for the collection and processing of massive bioacoustic data of sperm whales, detecting their basic communication units and language-like higher-level structures, and validating these models through interactive playback experiments. The technological capabilities developed by such an undertaking are likely to yield cross-applications and advancements in broader communities investigating non-human communication and animal behavioral research.

# Introduction

For centuries, humans have been fascinated by the idea of understanding the communication of animals (1). Animals live complex lives and often use signals to communicate with conspecifics for a variety of purposes throughout their daily routines; yet many have argued that their communication systems are not comparable, quantitatively or qualitatively, to that of the human languages. Human languages derive their expressive power from a number of distinctive structural features, including displacement, productivity, reflexivity, and recursion. Whether known non-human communication systems exhibit similarly rich structure — either of the same kind as human languages, or completely new — remains an open question.

Understanding language-like communication systems requires answering three key technical questions: First, by analogy to the phonetics and phonology of human languages, what are the *articulatory and perceptual building blocks* that can be reliably produced and recognized? Second, by analogy to the morphology and syntax of human languages, what are the *composition rules* according to which articulatory primitives can be knitted together? Third, by analogy to semantics in human languages, what are the interpretation rules that assign *meanings* to these building blocks? Finally, there may possibly be a *pragmatics* component, whereby meaning is additionally formed by *context* (2). While individual pieces of these questions have been asked about certain animal communication schemes, a general-purpose, automated, large-scale data-driven toolkit that can be applied to non-human communication is currently not available.

Animal communication researchers have conducted extensive studies of various species, including spiders (*e.g.* (3, 4)), pollinators (*e.g.* (5)), rodents (*e.g.* (6, 7)), birds (*e.g.* (8, 9)), primates (*e.g.* (2, 10–14)), and cetaceans (*e.g.* (15, 16)), showing that animal communication is multidimensional, involves diverse strategies, functions, and hierarchical components, and encompasses multiple modalities. Previous research efforts often focused on the mechanistic, computational, and structural aspects of animal communication systems. In human care, there have been several successful attempts of establishing a dialogue with birds (*e.g.* (17)) and primates through a shared, trained, anthropocentric lexicon or various media such as iconographic keyboards (*e.g.* (18)) or sign-language (*e.g.* (19). Due to the complexity of the environment and logistical challenges, such studies are often limited in sample size, continuity, and duration.

A comparatively long list of skills required for language learning in humans has been demonstrated among cetaceans (whales, dolphins, and porpoises), who share many social characteristics that are strikingly similar to our own. Whales and dolphins are among a few animals capable of vocal production learning (the ability to copy novel sounds as well as to vary those to produce individually distinctive repertoires) in addition to some birds, bats, pinnipeds, and elephants (20–22). Of those, only a few species, including parrots and dolphins appear to use arbitrary, learned signals to label objects or conspecifics in their communities in the wild(20, 23–26). Dolphins can use learned vocal labels to refer to and address each other when they meet at sea (27, 28). This sort of vocal recognition system mediates highly dynamic societies among cetaceans, which involve social relationships lasting decades as well as regular interaction with strangers (e.g. (29–33).

At the same time, cetaceans provide a dramatic contrast in their ecology and environment compared to terrestrial animals (34). The logistical and technological difficulties related to the observation of marine life are one of the reasons why relatively little is known about many of the toothed whales (Odontocetes). For example, it was not until 1957 that it was even noted that sperm whales (*Physeter macrocephalus)* produce sound (35) and only in the 1970s came the first understanding that they use sound for communication (36). Among all odontocetes species, *P. macrocephalus* stands out as an "animal of extremes" (37, 38). Sperm whales are the largest of the toothed whales, amongst the deepest divers, and have a circumglobal distribution (37). They can be both ocean nomads and small island specialists whose homes are both thousands of kilometers across and thousands of meters deep (39). Sperm whales' immense nose, the origin of their biological name ('macrocephalus' translates as 'large head'), houses the world's most powerful biological sonar system, which in turn is controlled by the world's largest brain, six times heavier than a human one (40–42) and with large cerebral hemispheres and spindle neurons (43–47). These cerebral structures might be indicative of complex cognition and higher-level functions put by sperm whales to task in both their rich social lives and the complex communication system.

While significant advances have been made with smaller cetaceans in human care using the advantage of a controlled experimental environment, results obtained in captivity always require replication in the natural habitat and raise ethical concerns. For these reasons, one should prefer to conduct studies in the open ocean. Yet, an often overlooked reality when comparing marine species with their terrestrial counterparts is that the ocean is large across all dimensions. Many whales cover thousands of kilometers (*e.g.* (48) and some are thought to live longer than a hundred years (49). Compared to terrestrial counterparts, marine species also experience substantially greater environmental variation over periods of months or longer (34), creating a situation in which social learning is favored over individual learning or genetic determination of behavior. Together with the fact that many cetaceans live in stable social groups with prolonged parental care, the opportunities for cultural transmission of information and traditional behaviors are high with traits being passed consistently within social groups, but less often between them. As a result, several cetacean species exhibit high levels of behavioral variation between social groups, much of which is thought to be due to social learning. The marine environment renders chemical or physical signals relatively ineffective, and whales rely on acoustics as their primary mode of communication. Most of their communication is thus likely to be captured by a single modality, making it an easier subject to study.

While multiple efforts in past decades to analyze non-human communication have brought a significant new understanding of various animal species and the structure and function of their signals, we still largely lack a functional understanding of non-human communication systems. In retrospect, we can conclude that critical understanding was acquired slowly across long periods of time invested with specific communities of a limited number of species and a modestly-sized amount of data for each. This is contrasted with the rapid growth of technologies that allow one to collect and process huge amounts of data. One such technology is Machine Learning (ML), in particular, deep learning (50) that has had a dramatic impact in natural language processing (NLP). Over the past decade, advances in machine learning have provided new powerful tools to manipulate language, making it now possible to construct unsupervised human language models capable of accurately capturing numerous aspects of phonetics and phonology, syntax, sentence structure, and semantics. Today's state-of-the-art NLP tools can segment low-level phonetic information into phonemes, morphemes, and words (51), turn word sequences into grammars (52–54), and ground words in visual perception and action (55–57). These NLP tools can be transferred from natural language to non-human vocalizations in order to identify patterns that would be difficult to discover with a manual analysis. However, interpretable models of both human language and animal communication rely on formal approaches and theoretical insights (2, 58). ML outputs are thus primarily a tool to constrain hypothesis space based to build formal and interpretable descriptions of the sperm whale communication. Combining key concepts from machine learning and linguistic theory could thus substantially advance the study of non-human communication and, more broadly, bring a data-centric paradigm shift to the study of animals.

The success of ML methods in NLP applications is related to the availability of large-scale datasets. The effort of creating a biological dataset in a format, level of detail, scale, and time span amenable to ML-based analysis is capital intensive and necessitates a multidisciplinary expertise to develop, deploy, and maintain specialized hardware to collect acoustic and behavioral signals, as well as software to process and analyze them, develop linguistic models that reveal the structure of animal communication and ground it in behavior, and finally perform playback experiments and attempt bidirectional communication (**Figure 1**). At the same time, the success of such an endeavor could potentially yield cross-applications and advancements in broader communities investigating non-human communication and animal behavioral research.

In this paper, we lay out a scientific and technological roadmap towards understanding sperm whale communications, which builds upon recent advances at the intersection of robotics, machine learning, natural language processing, marine biology, and linguistics. The goal is to provide a blueprint for other non-human communication projects, where various components can be modified to account for differing communication systems. We describe the current state of knowledge on sperm whale communication and outline the key ingredients of the collection and processing of massive bioacoustic data from sperm whales, detecting their basic communication units, language-like higher-level features, and discourse structure. We discuss experiments required to validate linguistic models and attribute meaning to communication units, and conclude with perspectives about the future progress in the field.

## Background

Sperm whales are born into tightly-knit matrilineal families within which females (who are not always related) and their offspring make group decisions when traveling (59), finding food and foraging together (37). Family members communally defend and raise their offspring, including nursing each others' calves (37, 60, 61). Some families join up for hours to a few days to form 'groups' with evidence of decade-long associations (32). On a higher level, sperm whales form clans of up to hundreds to tens of thousands of individual whales and exhibit diversity in movement patterns, habitat use, diving synchronization, foraging tactics, and diet; these differences appear to impact survival (62–65). Sperm whale clans coexist in overlapping ranges but remain socially segregated, despite not being genetically distinct communities (66).

**Acoustic communication of sperm whales**

Despite its present-day use for communication, the sperm whales' remarkable bioacoustic system (see **Figure 2A**) evolved as a sensory device for echolocation allowing the whales to find prey and navigate in the darkness of the deep ocean (40, 67). Each short, highly directional, broadband echolocation click has a multi-pulse structure with an intense first pulse followed by a few additional pulses of decaying amplitude (see **Figure 2B**). The multi-pulsed click is the result of the reverberation of the initial pulse in the whale's *spermaceti organ* within its nose (41, 68).

Whale communication utilizes short (<2 seconds) bursts of clicks produced in stereotyped patterns that can be classified into recognizable types termed *codas* (36, 69) (see **Figure 2B**). Distinct vocal sperm whale dialects have been documented in the Pacific, Indian, and Atlantic oceans (31, 70–74). Each distinct socially learned clan dialect contains at least 20 different coda types. A typical coda is made up of 2—40 broadband omnidirectional clicks. Codas are produced most prolifically during longer periods of intense socialization near the surface when sperm whales are in close contact, at the onset of deep foraging dives, as well as during ascent when approaching the surface, but not when at depth foraging (37, 75). Recent insights into the coda repertoires used by individuals and groups of whales have suggested that specific codas encode varying levels of social recognition to mediate the animals' complex multi-level societies (70). Codas appear to be rich in information about the caller's identity and there is some understanding of the diversity of coda types and the patterns of variation in their usage. Yet, the communicative function of particular codas themselves is still largely a mystery.

Codas are exchanged in duet-like sequences between two or more sperm whales. There is apparent turn-taking with whales responding within two seconds of each other, often overlapping and matching identical calls (76). These exchanges occur across spatial scales ranging from meters to kilometers, suggesting that they function both between whales immediately together and those farther apart. Individuals within a family share a natal dialect of at least 10 coda types, despite there being some variation in individual production repertoires (70, 77). Calves take at least two years to produce recognizable coda types and appear to 'babble' in producing a larger number of call types prior to narrowing their usage to the types produced by their natal family (70).

**Machine learning for bioacoustic signal processing and analysis**

The time and capital investment as well as technical and logistical challenges connected to collecting high-quality field audio recordings and subsequently manually annotating and analyzing them have been a key factor to the relatively slow pace in the study of sperm whale communication. Given these challenges, the development of improved computational techniques for automatic processing, annotation, and analysis of information content and communicative intent of whale vocalizations is a crucial step for future progress in the field. Machine learning and natural language processing tools provide great potential in addressing these challenges. Encouraging results in this direction were shown by (78), who used ML methods to automatically detect clicks in whale vocalization recordings, distinguish between echolocation and communication clicks, and classify codas into clans and individuals, achieving accuracy similar to previous highly time-consuming manual annotations and older generation statistical techniques.

Today's ML systems used in natural language processing applications are predominantly based on *deep representation learning*: input signals (e.g. sentences or audio waveforms) are encoded as high-dimensional feature vectors by an artificial neural network; these features are then decoded by another neural network into predictions for a downstream task (e.g. text classification or machine translation). The encoder network can be trained without labels via "self-supervision," typically to produce representations that make it possible to reconstruct parts of the input that have been hidden or corrupted. This apparently simple task requires a deep understanding of the structure of the language and creates a rich language representation that can be used for a plethora of tasks, including automated grammar induction (52–54) and machine translation without parallel data (79).

However, a key characteristic of this self-supervision process is its reliance on massive collections of data, with state-of-the-art language models such as GPT-3 (80) using over $10^{11}$ data points for training. While it is hard to make an exact analogy between tokens in human languages and whale vocalizations, for comparison, the Dominica Sperm Whale Project (DSWP) dataset used by (78) contained less than $10^4$ coda clicks (**Figure 3**). DWSP has hosted a longitudinal study since 2005. It is thus apparent that one of the key challenges towards the analysis of sperm whale (and more broadly, animal) communications using modern deep learning techniques is the need for datasets comparable in size to those used in NLP. Secondly, human linguistic corpora are easier to deal with because they are typically *pre-analyzed* (i.e., already presented in the form of words or letter), whereas in bioacoustic communication data the relevant units must be inferred bottom-up.

# Recording and Processing: Building the sperm whale longitudinal dataset

**Data acquisition**

Large scale data collection over lengthy timespans (years of recordings and observation) requires the use of autonomous and semi-autonomous assets that continuously operate on, around, and above the whales (**Figure 4**). Multiple technologies available today can be utilized for purposes including localization of groups of sperm whales, time- and location-stamped audio recording, and collection of other data such as ocean conditions and video capturing of whales' behavior. Assets coming in contact with whales should be designed with non-invasive technology (81) in order to minimize disturbance to animals, which in turn would provide more reliable data and also be more respectful to the study subject. Finally, the location for data collection should ideally have a known large resident sperm whale population.

*Tethered buoy arrays* (**Figure 4b**) are a typical setup utilized for background recording of bioacoustic signals. Such installations usually comprise an array of sensors mounted at intervals of several hundred meters from the surface to the depth at which sperm whales are known to hunt, approximately 1200m. The use of multiple sensors on each mooring and multiple moorings should allow to localize the whales and track their movements. The advantage of such arrays is their reliability and capability to record signals continuously from a broad area in the ocean.

*Tags* (**Figure 4c**), or recording devices attached to whales have historically provided the most detailed insight into their daily activities and interactions (Johnson and Tyack, 2003). There are currently several designs of animal-borne recording devices that use suction to delicately attach to the whales and record not only the whale acoustics but also pressure, temperature, movement and orientation. A critical current limitation of tags is onboard energy storage and memory as well as the effectiveness of their adhesion mechanisms. Bioinspired suction-based adhesion mechanisms inspired by carangiform fish (82, 83) and cephalopod tentacles hold the promise of achieving working times on the order of several days to potentially weeks. Fused with the sensor array data, the recordings from tags also allow the identification of whales in multi-party discourses and associating behavior patterns with background recordings of the hydrophone/static sensor arrays.

*Aquatic drones* (**Figure 4d & e**)*:* Free-swimming and passively floating aquatic drones allow obtaining audio and video recordings from multiple animals simultaneously to observe behaviors and conversations within a group of whales near the surface. There is a wide spectrum of potential solutions from simple drifters to self-propelled robots capable of autonomous navigation, including numerous existing platforms that can be loosely categorized as active, submarine-like bodies or semi-passive "gliders". For self-propelled drones, small, short-range, bioinspired designs (84, 85) hold the potential to operate in close proximity to a group of whales with minimal disruption.

*Aerial drones* (**Figure 4a**)*:* Hybrid aerial/aquatic drones are capable of surveying areas to monitor the whale population, and providing "just-in-time" deployment of hydrophones and possibly also deploying and recovering tags.

**Data processing**

Given the large magnitude of data, a key step is to build appropriate data storage and processing infrastructure, including automated ML pipelines (maintainable and reusable across multiple data collecting devices) that will replace the annotation currently done largely by hand by marine biologists. ML-based methods are already being used for click detection and classification (78); such methods are potentially scalable to large datasets containing years of recording that would otherwise be beyond reach with previous manual approaches.

By aggregating and synchronizing the bioacoustic, behavioral, and environmental signals from multiple assets (**Figure 4**), it is possible to localize the whales and continuously track them over time. The resulting dataset, a sort of "social network" of whales, will provide longitudinal information about the individual whales' behavior and their communications (86–88) and will be a crucial asset for subsequent machine learning.

# Decoding and Encoding: Building the sperm whale communication model

In human languages, there has been substantial recent progress in automated processing and unsupervised discovery of linguistic structure, including acoustic representation learning (89, 90), text generation (80), induction of phrase structure grammars (52–54), unsupervised translation (79, 91), and grounding of language in perception and action (92, 93), based on large-scale datasets. Similar tools could be applied to automatically identify structure in whale vocalizations.

**Phonetics and phonology: basic acoustic building blocks**

One of the most striking features of human language is its discrete structure. While the sound production apparatus and the acoustic speech stream are fundamentally continuous (humans can modulate pitch, volume, tongue position, etc. continuously), all human spoken languages partition this space into a discrete set of vowels, consonants, and tones (94, 95). Even though these discrete mental representations of sounds (*phonemes)* do not carry meaning, they form the building blocks from which larger meaning-carrying components are built. The distribution of phonemes in human languages is governed by a set of rules (phonotactic and phonological) that have also been identified, in a similar but simpler form, in vocalizations of non-human species, such as birds (58).

Previous research has conjectured that sperm whale communication is also built from a set of discrete units. Codas -- prototypical sequences of clicks with fixed relative inter-click interval structure -- have been identified as such fundamental communicative discrete units (69, 76). However, a plethora of questions remain: are codas distinguished only by the absolute inter-click intervals, as suggested by the current literature (*e.g.* (70, 96)? Do spectral features of coda clicks carry information? Does the frequency of individual clicks in codas carry meaning? What are the distributional restrictions (equivalents of phonotactic rules) governing codas and how can they be formalized? Can we find equivalents of phonological computation in sperm whale vocalizations and what type of formal grammar best describes their vocalizations (97)?

Identifying the fundamental units in whale vocalizations resembles spoken term discovery in human speech processing (89), which has been addressed with a variety of unsupervised learning techniques (98–101). These techniques use raw speech to automatically identify discrete clusters that represent repeated motifs -- thereby finding structure inherent in the data via *compression*. Such techniques are already effective at automatically identifying words from raw audio of human languages (90, 98, 99, 101–103).

Deep learning models for unsupervised discovery of meaningful units trained on human speech can readily be evaluated, as independent identification of meaningful units in speech is almost always available. However, in the case of sperm whale vocalizations, validation is substantially more challenging and necessitates the use of behavioral data and playback experiments. Unsupervised learning is most effective when applied to large and diverse datasets (applications in human speech perform best with hundreds to thousands of hours of recordings (104), highlighting the need for a large-scale bioacoustic data collection.

**Morphology and syntax: grammatical structure of communication**

The capacity to construct complex words and sentences from simpler parts according to regular rules is one of the hallmarks of human language. While compositional codes appear in some animal communication systems (e.g. the waggle dance in honeybees composes independent distance and orientation factors (105), and Campbell's monkeys use affixation to alter alarm call meaning (12), no known animal communication system appears to feature more complex structure-building operations like recursion, a central feature of almost all human languages. According to current knowledge, animal systems that have a semantics (e.g. primate calls and gestures or bird calls) appear to have a simple syntax; on the other hand, systems that have a somewhat sophisticated syntax (e.g. birdsongs, see (58) are not associated with a compositional semantics.

The complexity and duration of whale vocalizations suggest that they are at least in principle capable of exhibiting a more complex grammar, and it is thus important to attempt to find the rules of this grammar if they exist. In human languages, recent advances in NLP methods for *unsupervised grammar induction* (52–54) have shown the possibility of accurately recovering dependency and phrase structure grammars from a collection of sentences. Applying such techniques to the discretized "basic unit" sequences of whale communications should allow generating hypotheses about higher-level hierarchical structures across codas -- the *syntax* of whale vocalization. As with the representation learning approaches for identifying basic units, large datasets are crucial for this effort: since any given sequence can be explained by many different candidate grammars, many sequences are necessary to adequately constrain the hypothesis space.

**Semantics: inferring meaning**

Identifying short-term and long-term structure of vocalizations is a prerequisite to the key question: what do these vocalizations *mean*? The first step towards this goal is to identify the smallest meaning-carrying units, analogous to *morphemes* in human languages. It is known that individual codas carry information about the individual, family, and clan identity (70, 96, 106), but the function of many codas, as well as their internal variability in structure and individual clicks, remains unexplained. It is imperative that the collected data used for machine learning captures this richer context of whale vocalizations, enabling the grounding of a wider set of morphemes and candidate meanings.

The grounding of minimal units ("morphemes"), together with identified hierarchical structures allows to search for *interpretation rules* -- associations of complex behaviors with long sequences of clicks via an explicit bottom-up process. A number of compositional semantic models in the NLP community are capable of learning mappings between morpheme sequences and continuous groundings (107, 108). Currently existing whale bioacoustic datasets are likely too small for this purpose (see **Figure 3**), hence the need for acquiring a significantly larger and more detailed dataset. Finally, modeling composition should allow building a richer model of communicative intents, and ultimately to perform *interventional* studies in the form of playback experiments.

**Discourse and Social Communication**

Communication (whether human or non-human) occurs in a social context: speakers reason about interlocutors' beliefs and intentions (109), explicitly signal the beginning and end of their conversational turns (110), and adapt both the style and content of their messages to their audience (111). The complex, multi-party nature of sperm whale vocalization, and especially the presence of vocal learning and chorusing behaviors with no obvious analogue in human communication (76, 112, 113), suggests that this *discourse*-level structure is as important as the utterance-level structure for understanding whale communication.

Characterizing whales' *conversational protocols*, the rules that govern which individuals vocalize at what times, is key to understanding their discourse. Diverse communication protocols can be found across the animal kingdom -- including uncoupled responding after a pause, chorusing in alternation and chorusing synchronously -- and each of these evolved protocols has been found to provide distinctive advantages for competitive or cooperative reproductive advantage, food advantage, and territorial defense (114). Variants of all these protocols have been observed in sperm whales (76) and it is necessary to understand the roles that each of them plays vis-a-vis clan structure and group decision-making.

The understanding of conversational protocols is also a prerequisite to building *predictive models of conversations* (analogous to *language models* and *chatbots* for human-generated speech and text (80, 115, 116) capable of generating probable vocalizations given a conversation history, whale identities, and behavioral and environmental context. These models can be made controllable and capable of continuing vocalizations to express specific communicative intents (using inferred meanings for vocalizations in historical data) and will enable interactive playback studies.

**Redundancy and Fault Tolerance of Communication**

Most forms of communication rely on the capacity to successfully transmit and receive a sequence of some basic units. In instances of imperfect acoustic channels with significant background noise, fault tolerance mechanisms are sometimes built into the communication system at different levels. In the animal kingdom, multiple fault tolerance mechanisms are known, that exploit varying modalities to backup communication signals (117); or adapt the communication units to noise conditions (118). Sperm whales, for example, have been shown to repeat vocalizations, including overlapping and matching codas (76), a characteristic that might suggest redundancy mechanisms at the level of basic units and discourse. As studies venture into this area, it is important that such variations are distinguished from dialectal and individual variations, which can be detected using e.g. compression-based techniques (119).

**Language acquisition**

All human infants undergo similar stages during acquisition of language in the first years of life, regardless of the language in their environment. For example, the babbling period during which language-acquiring infants produce and repeat basic syllables (such as [da] or [ba]) or reduced handshapes and movements in sign languages (120) is a well-documented developmental stage during the first 6-13 months (121). Another well-documented concept in language acquisition is the critical period: if children are deprived of primary linguistic inputs in their first years, acquisition is not complete, often resulting in severe linguistic impairments (Friedmann 2015). The study of the developmental stages in language acquisition has yielded insights into how humans learn to discretize the acoustic speech stream into mental units, analyze meaning, and in turn produce language. In human language, for example, syllables that are produced first during language acquisition (e.g. [ma] or [ba]) are also most common in the world's languages, most stable, and easiest to produce. Similarly, morphological and syntactic constructions that are acquired first are the most basic (122).

There are currently several known parallels in the developmental stages between human language and animal communication. Acquisition of birdsong in some species, for example, involves the presence of babbling as well as the critical period (123). These parallels likely stem from common neural and genetic mechanisms behind human speech and animal vocalizations (124). However, in cetacean research, existing data on the vocalizations of non-adult whales in their natural setting are limited. Continuous and longitudinal data acquisition capabilities are required to record vocalizations of calf-mother pairs and collect behavioral data on their interactions as calves mature. Such data will provide insights into the order of acquisition of coda types, leading to insights into the articulatory effort of the vocalization as well as identification of the most basic structural building blocks and their functions.

## Playback-based Validation

Playbacks are the experimental presentation of stimuli to animals, traditionally used to investigate their behavioral, cognitive, or psychophysiological responses (125). Playbacks in relation to animal communication can be categorized based on (i) the stimulus type (such as responses to conspecific or heterospecific signals (e.g. (126, 127) or anthropogenic noise (e.g., sonar behavioral response studies, reviewed in (128) and (ii) the collected data (such as response calls or behavior). While playback validation is a common technique used to study the vocalizations of terrestrial animals including birds (129), primates (130), and elephants (131, 132) that has proven successful in both grounding the functional use of calls as well as building understanding of the physiological and cognitive abilities of these animals In cetacean research, the vast majority of playback experiments have focused on the functional use of calls for social identity. It was shown this way, for example, that bottlenose dolphins use vocal labels to address one another (24).

For any vocal recognition system to function in this way, it must meet the following three criteria: first, there must be calls that vary enough and/or are sufficiently stereotyped to provide identity information for individuals or groups; second, listeners must be able to distinguish between these calls and hold a shared meaning/function for the calls; and third, listeners must then respond differently to those calls based on the identity of the signaller and their interaction history with them.

The divide between playbacks *in-situ* at sea and the captive experiments is partly a result of a separation in focus: captive studies primarily examined the auditory capacity and cognitive responses of the animals, while wild studies mainly looked at response to conspecific calls. Another major separating factor is the logistical and technological limitations of performing playback studies at sea: studying animal communication in the wild within a natural social and behavioral context is significantly harder than in controlled settings in captivity. However, despite their complexity, wild playback experiments increase functional validity by avoiding the disturbance of species-typical social groups and daily behavioral routines (133).

The inherent challenges of conducting playback experiments for the purpose of grounding hypotheses of any animal communication model fall under three general questions (see (134):

1) Do we know *what to playback*? Formalizing hypotheses requires a detailed understanding of both the signals being produced and the social/behavioral context in which they are used, and must be preceded by addressing core phonological, syntactical, and semantic questions using language models to better build appropriate playback stimuli to underlie grounding experiments within behavioral contexts.

2) Do stimuli *replicate biological signals*? Playback signals must adequately replicate the parameters of the natural signals themselves, avoid pseudo-replication with a sufficiently large sample, and reduce the logistical and perceptual limitations of conducting field playbacks from boats with researchers present. This requires developing novel playback technology based on autonomous aquatic robots, which remove the vessel from the experiment.

3) Can we *recognize a response*? The ability to detect and identify behavioral responses to the playback stimuli requires a baseline understanding of the variation in behavior in the wild from observational studies. This is perhaps the biggest challenge as it requires both an understanding of what whales do, but also what we expect them to do in response to our playbacks.

While cetacean playbacks have similar interpretation challenges as terrestrial studies, they are logistically more challenging and mainly technologically-limited. The purpose of playback experiments in Project CETI is two-fold. First, and more typical, is the use of playbacks to ground semantic hypotheses and test purported syntax based on hypotheses generated from language models. The second use case, which can be viewed as an evolution of the first one, is more speculative but potentially offering an opportunity to make significant advancement in field-based, interactive-playback among whales. There is currently rapid innovation of interactive playbacks in which researchers are able to more rapidly reply to animals' communication in the wild. This is particularly evident in bird song research (135). Technological development of tools in this area in some ways mirrors advances with increasingly common NLP-based interactive voice assistants and chatbots, which are intended to listen, detect, and appropriately reply in context to their human users.

## Conclusion

Recent advances in machine learning developed for the analysis of human-generated signals and broadly used in industry now make it possible to obtain unprecedented insights into the structure and meaning of non-human species communication. Such methods, when applied to purposely built datasets, are likely to bring a shift in perspective in deciphering animal communication in their natural settings. Achieving this ambitious goal requires an orchestrated effort and expertise from multiple disciplines across the scientific community. A prerequisite for this to happen is an open source and data sharing culture that has allowed the machine learning research community to flourish over the past decade.

Previous large-scale and collaborative efforts have been successful in yielding substantial steps forward in the understanding of natural systems. Past collaborative projects (in particular, in genetics and astrophysics) turned out to be influential "not because they answer any single question but because they enable investigation of continuously arising new questions from the same data-rich sources" (136), and their impact resulted from providing the technological foundations as well as findings and advancements along the journey.

Beyond advancing our understanding of natural communication systems, we see these efforts leading to tool sets that can be utilized in a diversity of fields. A large-scale, interdisciplinary and integrated study of cetacean communication will also advance the design of underwater acoustic sensors, gentle robotics, processing complex bioacoustic signals, and machine learning for language modeling. Collective advances in this area hold the potential to open new frontiers in interspecies communication and can lead to a deeper appreciation and understanding of the complexity and diversity of communication in the natural world.

## Acknowledgments

Project CETI (Cetacean Translation Initiative), a nonprofit organization, applying advanced machine learning and non-invasive robotics to listen to and translate the communication of whales is funded by grants from Dalio Philanthropies and Ocean X; Sea Grape Foundation; Rosamund Zander and Hansjorg Wyss, and others, through the Audacious Project: a collaborative funding initiative housed at TED; as well as National Geographic Society Grant (No. NGS-72337T-20). We thank Jane Lipson and Pietro Lio' for helpful comments on the manuscript.

# References

1. Aristotle., R. Cresswell, J. G. Schneider, *Aristotle's History of animals : In ten books*, v. 7 (H.G. Bohn, 1862).
2. P. Schlenker, *et al.*, Formal monkey linguistics. *Theor. Linguist.* **42**, 1–90 (2016).
3. E. A. Hebets, C. J. Vink, L. Sullivan-Beckers, M. F. Rosenthal, The dominance of seismic signaling and selection for signal complexity in Schizocosa multimodal courtship displays. *Behav. Ecol. Sociobiol.* **67**, 1483–1498 (2013).
4. D. O. Elias, W. P. Maddison, C. Peckmezian, M. B. Girard, A. C. Mason, Orchestrating the score: complex multimodal courtship in the Habronattus coecatus group of Habronattus jumping spiders (Araneae: Salticidae). *Biol. J. Linn. Soc. Lond.* **105**, 522–547 (2012).
5. I. G. Kulahci, A. Dornhaus, D. R. Papaj, Multimodal signals enhance decision making in foraging bumble-bees. *Proc. R. Soc. B Biol. Sci.* **275**, 797–802 (2008).
6. S. H. Ackers, C. N. Slobodchikoff, Communication of Stimulus Size and Shape in Alarm Calls of Gunnison's Prairie Dogs, Cynomys gunnisoni. *Ethology* **105**, 149–162 (1999).
7. C. N. Slobodchikoff, A. Paseka, J. L. Verdolin, Prairie dog alarm calls encode labels about predator colors. *Anim. Cogn.* **12**, 435–439 (2009).
8. M. C. Baker, "Bird Song Research: The Past 100 Years" (2001).
9. M. Griesser, D. Wheatcroft, T. N. Suzuki, From bird calls to human language: exploring the evolutionary drivers of compositional syntax. *Curr. Opin. Behav. Sci.* **21**, 6–12 (2018).
10. C. B. Jones, T. E. Van Cantfort, Multimodal communication by male mantled howler monkeys (Alouatta palliata) in sexual contexts: A descriptive analysis. *Folia Primatol.* **78**, 166–185 (2007).
11. D. A. Leavens, Animal Cognition: Multimodal Tactics of Orangutan Communication.

*Curr. Biol.* **17**, R762--R764 (2007).

12. K. Ouattara, A. Lemasson, K. Zuberbühler, Campbell's monkeys concatenate vocalizations into context-specific call sequences. *Proc. Natl. Acad. Sci. U. S. A.* **106**, 22026–22031 (2009).
13. E. Clarke, U. H. Reichard, K. Zuberbühler, The Syntax and Meaning of Wild Gibbon Songs. *PLoS One* **1**, e73 (2006).
14. R. M. Seyfarth, D. L. Cheney, P. Marler, Vervet monkey alarm calls: Semantic communication in a free-ranging primate. *Anim. Behav.* **28**, 1070–1094 (1980).
15. V. M. Janik, L. S. Sayigh, Communication in bottlenose dolphins: 50 years of signature whistle research. *J. Comp. Physiol. A Neuroethol. Sensory, Neural, Behav. Physiol.* **199** (2013).
16. V. M. Janik, *Cetacean vocal learning and communication* (Elsevier Ltd, 2014).
17. I. M. Pepperberg, Cognition in an African gray parrot (Psittacus erithacus): Further evidence for comprehension of categories and labels. *J. Comp. Psychol.* **104**, 41–52 (1990).
18. S. Savage-Rumbaugh, D. M. Rumbaugh, K. McDonald, Language learning in two species of apes. *Neurosci. Biobehav. Rev.* **9**, 653–665 (1985).
19. F. G. Patterson, The gestures of a gorilla: Language acquisition in another pongid. *Brain Lang.* **5**, 72–97 (1978).
20. V. M. Janik, P. J. B. Slater, Context-specific use suggests that bottlenose dolphin signature whistles are cohesion calls. *Anim. Behav.* **56**, 829–838 (1998).
21. V. M. Janik, P. J. B. Slater, Vocal Learning in Mammals. *Adv. Study Behav.*, 59–99 (1997).
22. J. H. Poole, P. L. Tyack, A. S. Stoeger-Horwath, S. Watwood, Elephants are capable of vocal learning. *Nature* **434**, 455–456 (2005).
23. T. J. S. Balsby, J. W. Bradbury, Vocal matching by orange-fronted conures (Aratinga canicularis). *Behav. Process.* **82**, 133–139 (2009).
24. S. L. King, L. S. Sayigh, R. S. Wells, W. Fellner, V. M. Janik, Vocal copying of individually distinctive signature whistles in bottlenose dolphins. *Proc. Biol. Sci.* **280**, 20130053 (2013).
25. P. L. Tyack, L. S. Sayigh, Vocal learning in cetaceans. *Soc. Influ. Vocal Dev.*, 208–233 (1997).
26. R. Wanker, Y. Sugama, S. Prinage, Vocal labelling of family members in spectacled parrotlets, Forpus conspicillatus. *Anim. Behav.* **70**, 111–118 (2005).
27. S. L. King, V. M. Janik, Bottlenose dolphins can use learned vocal labels to address each other. *Proc. Natl. Acad. Sci. U. S. A.* **110**, 13216–13221 (2013).

28. N. J. Quick, V. M. Janik, Bottlenose dolphins exchange signature whistles when meeting at sea. *Proc. R. Soc. B Biol. Sci.* **279**, 2539–2545 (2012).

29. P. Tyack, Population biology, social behavior and communication in whales and dolphins. *Trends Ecol. Evol.* **1**, 144–150 (1986).

30. J. N. Bruck, Decades-long social memory in bottlenose dolphins. *Proc. Biol. Sci.* **280**, 20131726 (2013).

31. S. Gero, A. Bøttcher, H. Whitehead, P. T. Madsen, Socially segregated, sympatric sperm whale clans in the Atlantic Ocean. *R. Soc. Open Sci.* **3** (2016).

32. S. Gero, J. Gordon, H. Whitehead, Individualized social preferences and long-term social fidelity between social units of sperm whales. *Anim. Behav.* (2015) https:/doi.org/10.1016/j.anbehav.2015.01.008.

33. R. C. Connor, "Group Living in Whales and Dolphins" in *Cetacean Societies: Field Studies of Dolphins and Whales*, (2000).

34. J. H. Steele, A comparison of terrestrial and marine ecological systems. *Nature* **313**, 355–358 (1985).

35. L. V. Worthington, W. E. Schevill, Underwater sounds heard from sperm whales. *Nature* **4850**, 291 (1957).

36. W. A. Watkins, W. E. Schevill, Sperm whale codas. *J. Acoust. Soc. Am.* **62**, 1486–1490 (1977).

37. H. Whitehead, *Sperm whales: Social evolution in the ocean* (University of Chicago Press, 2003).

38. L. S. Weilgart, H. Whitehead, K. Payne, A colossal convergence - Sperm whales and elephants share similar life histories and social structures, which include social females and roving males. *Am. Sci.* **84:3** (1996).

39. M. Cantor, S. Gero, H. Whitehead, L. Rendell, "Sperm Whale: The Largest Toothed Creature on Earth" in *Ethology and Behavioral Ecology of Odontocetes*, B. Würsig, Ed. (Springer International Publishing, 2019), pp. 261–280.

40. J. A. Goldbogen, P. T. Madsen, The evolution of foraging capacity and gigantism in cetaceans. *J. Exp. Biol.* **221**, jeb166033 (2018).

41. B. Møhl, M. Wahlberg, P. T. Madsen, A. Heerfordt, A. Lund, The monopulsed nature of sperm whale clicks. *J. Acoust. Soc. Am.* **114**, 1143–1154 (2003).

42. L. Marino, A comparison of encephalization between odontocete cetaceans and anthropoid primates. *Brain Behav. Evol.* **51**, 230–238 (1998).

43. L. Marino, *et al.*, Cetaceans have complex brains for complex cognition. *PLoS Biol.* (2007) https:/doi.org/10.1371/journal.pbio.0050139.

44. L. Marino, Cetacean Brain Evolution: Multiplication Generates Complexity. *Int. J.*

*Comp. Psychol.* **17**, 1–16 (2004).

45. L. Marino, P. Brakes, M. P. Simmonds, "Brain structure and intelligence in cetaceans" in *Whales and Dolphins: Cognition, Culture, Conservation and Human Perceptions*, (EarthScan/Routledge New York, 2011), pp. 115–128.

46. C. Butti, C. C. Sherwood, A. Y. Hakeem, J. M. Allman, P. R. Hof, Total number and volume of Von Economo neurons in the cerebral cortex of cetaceans. *J. Comp. Neurol.* **515**, 243–259 (2009).

47. P. R. Hof, R. Chanis, L. Marino, Cortical complexity in cetacean brains. *Anat. Rec. Part A Discov. Mol. Cell. Evol. Biol.* **287A**, 1142–1152 (2005).

48. P. T. Stevick, *et al.*, A quarter of a world away: female humpback whale moves 10,000 km between breeding areas. *Biol. Lett.* **7**, 299–302 (2011).

49. I. Seim, *et al.*, The transcriptome of the bowhead whale Balaena mysticetus reveals adaptations of the longest-lived mammal. *Aging (Albany. NY).* **6**, 879–899 (2014).

50. Y. Lecun, Y. Bengio, G. Hinton, Deep learning. *Nature* (2015) https:/doi.org/10.1038/nature14539.

51. M. Elsner, S. Goldwater, J. Eisenstein, Bootstrapping a Unified Model of Lexical and Phonetic Acquisition in *Proceedings of the 50th Annual Meeting of the Association for Computational Linguistics (Volume 1: Long Papers)*, (Association for Computational Linguistics, 2012), pp. 184–193.

52. D. Klein, C. D. Manning, Corpus-based induction of syntactic structure: Models of dependency and constituency. *Proc. 42nd Annu. Meet. Assoc. Comput. Linguist.* (2004).

53. T. Naseem, H. Chen, R. Barzilay, M. Johnson, Using universal linguistic knowledge to guide grammar induction (2010).

54. Y. Kim, C. Dyer, A. M. Rush, Compound probabilistic context-free grammars for grammar induction. *arXiv Prepr. arXiv1906.10225* (2019).

55. S. Tellex, *et al.*, Understanding natural language commands for robotic navigation and mobile manipulation in *Proceedings of the Twenty-Fifth AAAI Conference on Artificial Intelligence*, AAAI'11., (AAAI Press, 2011), pp. 1507–1514.

56. A. Rohrbach, M. Rohrbach, R. Hu, T. Darrell, B. Schiele, Grounding of Textual Phrases in Images by Reconstruction in *Computer Vision -- ECCV 2016*, (Springer International Publishing, 2016), pp. 817–834.

57. J. Andreas, M. Rohrbach, T. Darrell, D. Klein, Neural module networks in *Proceedings of the IEEE Conference on Computer Vision and Pattern Recognition*, (2016), pp. 39–48.

58. R. C. Berwick, K. Okanoya, G. J. L. Beckers, J. J. Bolhuis, Songs to syntax: the

linguistics of birdsong. *Trends Cogn. Sci.* **15**, 113–121 (2011).

59. H. Whitehead, Consensus movements by groups of sperm whales. *Mar. Mammal Sci.* **32** (2016).

60. S. Gero, D. Engelhaupt, L. Rendell, H. Whitehead, Who cares? Between-group variation in alloparental caregiving in sperm whales. *Behav. Ecol.* **20**, 838–843 (2009).

61. S. Gero, J. Gordon, H. Whitehead, Calves as social hubs: dynamics of the social network within sperm whale units. *Proc. Biol. Sci.* **280**, 20131113 (2013).

62. H. Whitehead, L. Rendell, Movements, habitat use and feeding success of cultural clans of South Pacific sperm whales. *J. Anim. Ecol.* **73**, 190–196 (2004).

63. M. Cantor, H. Whitehead, How does social behavior differ among sperm whale clans? *Mar. Mammal Sci.* **31**, 1275–1290 (2015).

64. M. Marcoux, H. Whitehead, L. Rendell, Sperm whale feeding variation by location, year, social group and clan: Evidence from stable isotopes. *Mar. Ecol. Prog. Ser.* **333**, 309–314 (2007).

65. M. Marcoux, L. Rendell, H. Whitehead, Indications of fitness differences among vocal clans of sperm whales. *Behav. Ecol. Sociobiol.* **61**, 1093–1098 (2007).

66. L. Rendell, S. L. Mesnick, M. L. Dalebout, J. Burtenshaw, H. Whitehead, Can genetic differences explain vocal dialect variation in sperm whales, Physeter macrocephalus? *Behav. Genet.* **42**, 332–343 (2012).

67. P. Tønnesen, C. Oliveira, M. Johnson, P. T. Madsen, The long-range echo scene of the sperm whale biosonar. *Biol. Lett.* **16**, 20200134 (2020).

68. W. M. X. Zimmer, P. L. Tyack, M. P. Johnson, P. T. Madsen, Three-dimensional beam pattern of regular sperm whale clicks confirms bent-horn hypothesis. *J. Acoust. Soc. Am.* **117**, 1473–1485 (2005).

69. L. Weilgart, H. Whitehead, Group-specific dialects and geographical variation in coda repertoire in South Pacific sperm whales. *Behav. Ecol. Sociobiol.* (1997) https:/doi.org/10.1007/s002650050343.

70. S. Gero, H. Whitehead, L. Rendell, Individual, unit and vocal clan level identity cues in sperm whale codas. *R. Soc. Open Sci.* (2016) https:/doi.org/10.1098/rsos.150372.

71. L. E. Rendell, H. Whitehead, Vocal clans in sperm whales (Physeter macrocephalus). *Proc. R. Soc. B Biol. Sci.* (2003) https:/doi.org/10.1098/rspb.2002.2239.

72. M. Amano, A. Kourogi, K. Aoki, M. Yoshioka, K. Mori, Differences in sperm whale codas between two waters off Japan: possible geographic separation of vocal clans. *J. Mammal.* **95**, 169–175 (2014).

73. T. O. S. Amorim, *et al.*, Coda repertoire and vocal clans of sperm whales in the western Atlantic Ocean. *Deep Sea Res. Part I* **160**, 103254 (2020).

74. L. A. E. Huijser, *et al.*, Vocal repertoires and insights into social structure of sperm whales ( Physeter macrocephalus ) in Mauritius, southwestern Indian Ocean. *Mar. Mammal Sci.* **36**, 638–657 (2020).
75. S. L. Watwood, P. J. O. Miller, M. Johnson, P. T. Madsen, P. L. Tyack, Deep-diving foraging behaviour of sperm whales (Physeter macrocephalus). *J. Anim. Ecol.* **75**, 814–825 (2006).
76. T. M. Schulz, H. Whitehead, S. Gero, L. Rendell, Overlapping and matching of codas in vocal interactions between sperm whales: insights into communication function. *Anim. Behav.* **76**, 1977–1988 (2008).
77. T. M. Schulz, H. Whitehead, S. Gero, L. Rendell, Individual vocal production in a sperm whale (Physeter macrocephalus) social unit. *Mar. Mamm. Sci.* **27**, 149–166 (2011).
78. P. C. Bermant, M. M. Bronstein, R. J. Wood, S. Gero, D. F. Gruber, Deep Machine Learning Techniques for the Detection and Classification of Sperm Whale Bioacoustics. *Sci. Rep.* (2019) https:/doi.org/10.1038/s41598-019-48909-4.
79. G. Lample, M. Ott, A. Conneau, L. Denoyer, M. Ranzato, Phrase-Based & Neural Unsupervised Machine Translation (2018).
80. T. B. Brown, *et al.*, Language Models are Few-Shot Learners (2020).
81. M. Tessler, *et al.*, Ultra-gentle soft robotic fingers induce minimal transcriptomic response in a fragile marine animal. *Curr. Biol.* **30**, 157–158 (2020).
82. K. M. Gamel, A. M. Garner, B. E. Flammang, Bioinspired remora adhesive disc offers insight into evolution. *Bioinspir. Biomim.* **14**, 56014 (2019).
83. Y. Wang, *et al.*, A biorobotic adhesive disc for underwater hitchhiking inspired by the remora suckerfish. *Sci Robot* **2** (2017).
84. A. D. Marchese, C. D. Onal, D. Rus, Autonomous Soft Robotic Fish Capable of Escape Maneuvers Using Fluidic Elastomer Actuators. *Soft Robot.* **1**, 75–87 (2014).
85. R. K. Katzschmann, J. DelPreto, R. MacCurdy, D. Rus, Exploration of underwater life with an acoustically controlled soft robotic fish. *Sci Robot* **3** (2018).
86. D. R. Farine, H. Whitehead, Constructing, conducting and interpreting animal social network analysis. *J. Anim. Ecol.* **84**, 1144–1163 (2015).
87. P. Sah, J. D. Méndez, S. Bansal, A multi-species repository of social networks. *Sci Data* **6**, 44 (2019).
88. S. Sosa, D. M. P. Jacoby, M. Lihoreau, C. Sueur, Animal social networks: Towards an integrative framework embedding social interactions, space and time. *Methods Ecol. Evol.* **12**, 4–9 (2021).
89. H. Kamper, A. Jansen, S. King, S. Goldwater, Unsupervised lexical clustering of

speech segments using fixed-dimensional acoustic embeddings in *2014 IEEE Spoken Language Technology Workshop (SLT)*, (2014), pp. 100–105.

90. Y.-A. Chung, C.-C. Wu, C.-H. Shen, H.-Y. Lee, L.-S. Lee, Audio Word2Vec: Unsupervised Learning of Audio Segment Representations using Sequence-to-sequence Autoencoder (2016).

91. M. Artetxe, G. Labaka, E. Agirre, K. Cho, Unsupervised Neural Machine Translation (2017).

92. J. Lu, D. Batra, D. Parikh, S. Lee, ViLBERT: Pretraining Task-Agnostic Visiolinguistic Representations for Vision-and-Language Tasks (2019).

93. H. Shi, J. Mao, K. Gimpel, K. Livescu, Visually Grounded Neural Syntax Acquisition (2019).

94. B. H. Repp, "Categorical Perception: Issues, Methods, Findings" in *Speech and Language*, N. J. Lass, Ed. (Elsevier, 1984), pp. 243–335.

95. P. D. Eimas, E. R. Siqueland, P. Jusczyk, J. Vigorito, Speech perception in infants. *Science (80-. ).* **171**, 303–306 (1971).

96. R. Antunes, *et al.*, Individually distinctive acoustic features in sperm whale codas. *Anim. Behav.* **81**, 723–730 (2011).

97. N. Chomsky, Three models for the description of language. *IRE Trans. Inf. Theory* **2**, 113–124 (1956).

98. B. van Niekerk, L. Nortje, H. Kamper, Vector-quantized neural networks for acoustic unit discovery in the ZeroSpeech 2020 challenge in *Interspeech 2020*, (ISCA, 2020).

99. G. Beguš, CiwGAN and fiwGAN: Encoding information in acoustic data to model lexical learning with Generative Adversarial Networks (2020).

100. Y.-A. Chung, H. Tang, J. Glass, Vector-Quantized Autoregressive Predictive Coding (2020).

101. A. Baevski, S. Schneider, M. Auli, vq-wav2vec: Self-Supervised Learning of Discrete Speech Representations (2019).

102. J. Chorowski, R. J. Weiss, S. Bengio, A. van den Oord, Unsupervised speech representation learning using WaveNet autoencoders (2019).

103. R. Eloff, *et al.*, Unsupervised acoustic unit discovery for speech synthesis using discrete latent-variable neural networks (2019).

104. Y. A. Chung, J. Glass, Speech2Vec: A sequence-to-sequence framework for learning word embeddings from speech in *Proceedings of the Annual Conference of the International Speech Communication Association, INTERSPEECH*, (2018).

105. J. R. Glass, Towards unsupervised speech processing. 1–4 (2012).

106. C. Oliveira, *et al.*, Sperm whale codas may encode individuality as well as clan

identity. *J. Acoust. Soc. Am.* (2016) https:/doi.org/10.1121/1.4949478.

107. R. Socher, A. Karpathy, Q. V Le, C. D. Manning, A. Y. Ng, Grounded Compositional Semantics for Finding and Describing Images with Sentences. *Trans. Assoc. Comput. Linguist.* **2**, 207–218 (2014).

108. J. Andreas, A. Dragan, D. Klein, Translating Neuralese (2017).

109. H. P. Grice, "Logic and conversation" in *Speech Acts*, (Brill, 1975), pp. 41–58.

110. H. Sacks, E. A. Schegloff, G. Jefferson, A simplest systematics for the organization of turn-taking for conversation. *Language (Baltim).* **50**, 696–735 (1974).

111. H. Giles, N. Coupland, J. Coupland, Accommodation theory: Communication, context, and consequence. *Context. Accommod.*, 1–68 (1991).

112. A. D. Patel, Language, music, syntax and the brain. *Nat. Neurosci.* **6**, 674–681 (2003).

113. L. Weilgart, H. Whitehead, Coda communication by sperm whales (Physeter macrocephalus) off the Galápagos Islands. *Can. J. Zool.* **71**, 744–752 (1993).

114. A. Ravignani, D. L. Bowling, W. T. Fitch, Chorusing, synchrony, and the evolutionary functions of rhythm. *Front. Psychol.* **5**, 1118 (2014).

115. C. E. Shannon, Prediction and entropy of printed English. *Bell Syst. tech. j.* **30**, 50–64 (1951).

116. J. Gao, M. Galley, L. Li, Neural approaches to conversational AI. *Found. Trends Inf. Retr.* **13** (2019).

117. R. A. Johnstone, Multiple displays in animal communication:'backup signals' and "multiple messages." *Philos. Trans. R. Soc. B Biol. Sci.* **351** (1996).

118. S. E. LaZerte, H. Slabbekoorn, K. A. Otter, Learning to cope: vocal adjustment to urban noise is correlated with prior experience in black-capped chickadees. *Proc. Biol. Sci.* **283** (2016).

119. W. Oliveira Jr, E. Justino, L. S. Oliveira, Comparing compression models for authorship attribution. *Forensic Sci. Int.* **228**, 100–104 (2013).

120. L. A. Petitto, P. F. Marentette, Babbling in the manual mode: evidence for the ontogeny of language. *Science (80-. ).* **251**, 1493–1496 (1991).

121. M. K. Fagan, Mean Length of Utterance before words and grammar: longitudinal trends and developmental implications of infant vocalizations. *J. Child Lang.* **36**, 495–527 (2009).

122. S. Crain, R. Thornton, Syntax acquisition: Syntax acquisition. *Wiley Interdiscip. Rev. Cogn. Sci.* **3**, 185–203 (2012).

123. A. J. Doupe, P. K. Kuhl, Birdsong and human speech: common themes and mechanisms. *Annu. Rev. Neurosci.* **22**, 567–631 (1999).

124. J. J. Bolhuis, K. Okanoya, C. Scharff, Twitter evolution: converging mechanisms in birdsong and human speech. *Nat. Rev. Neurosci.* **11**, 747–759 (2010).
125. S. L. King, You talkin' to me? Interactive playback is a powerful yet underused tool in animal communication research. *Biol. Lett.* **11**, 20150403 (2015).
126. L. S. Sayigh, *et al.*, Individual recognition in wild bottlenose dolphins: a field test using playback experiments. *Anim. Behav.* **57**, 41–50 (1999).
127. F. Visser, *et al.*, Disturbance-specific social responses in long-finned pilot whales, Globicephala melas. *Sci. Rep.* **6**, 28641 (2016).
128. B. L. Southall, D. P. Nowacek, P. J. O. Miller, P. L. Tyack, Experimental field studies to measure behavioral responses of cetaceans to sonar. *Endanger. Species Res.* **31**, 293–315 (2016).
129. P. K. McGregor, *et al.*, Design of Playback Experiments: The Thornbridge Hall NATO ARW Consensus. *Play. Stud. Anim. Commun.*, 1–9 (1992).
130. J. Fischer, R. Noser, K. Hammerschmidt, Bioacoustic field research: a primer to acoustic analyses and playback experiments with primates. *Am. J. Primatol.* **75**, 643–663 (2013).
131. A. S. Stoeger, A. Baotic, Information content and acoustic structure of male African elephant social rumbles. *Sci. Rep.* **6**, 27585 (2016).
132. K. McComb, G. Shannon, K. N. Sayialel, C. Moss, Elephants can determine ethnicity, gender, and age from acoustic cues in human voices. *Proc. Natl. Acad. Sci. U. S. A.* **111**, 5433–5438 (2014).
133. K. A. Cronin, S. L. Jacobson, K. E. Bonnie, L. M. Hopper, Studying primate cognition in a social setting to improve validity and welfare: a literature review highlighting successful approaches. *PeerJ* **5**, e3649 (2017).
134. V. B. Deecke, Studying marine mammal cognition in the wild: a review of four decades of playback experiments. *Aquat. Mamm.* **32**, 461–482 (2006).
135. T. Dabelsteen, P. K. McGregor, 22. Dynamic Acoustic Communication and Interactive Playback. *Ecol. Evol. Acoust. Commun. Birds*, 398–408 (2020).
136. L. F. Abbott, *et al.*, The mind of a mouse. *Cell* **182**, 1372–1376 (2020).

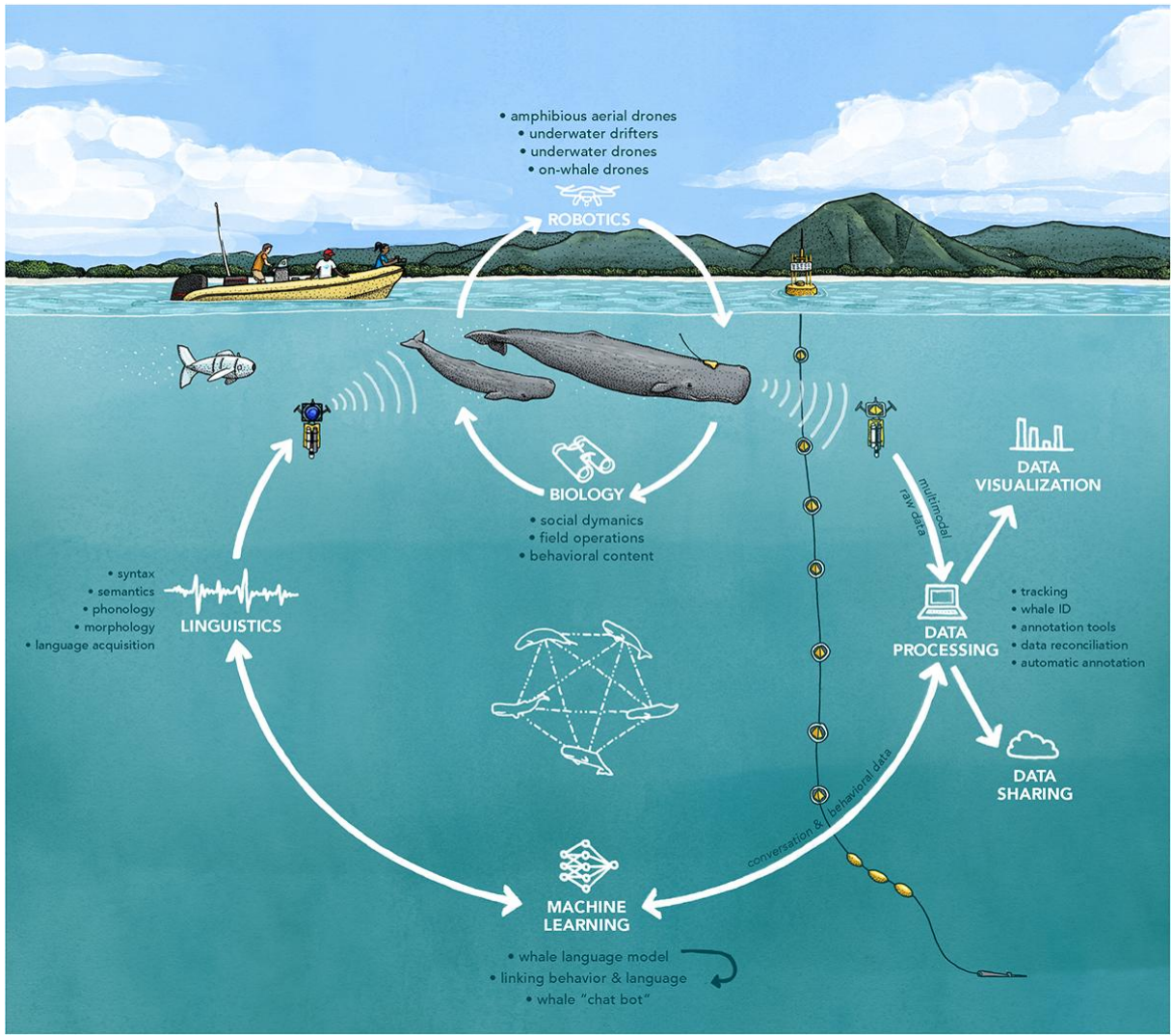

***Figure 1:*** *An interdisciplinary approach to sperm whale communication that integrates biology, robotics, machine learning, and linguistics expertise, and comprise the following key steps.* ***Record:*** *collect large-scale longitudinal multi-modal dataset of whale communication and behavioral data from a variety of sensors.* ***Process:*** *reconcile and process the multi-sensor data.* ***Decode:*** *using machine learning techniques, create a model of whale communication, characterize its structure, and link it to behavior.* ***Encode & Playback:*** *conduct interactive playback experiments and refine the whale language model.* Illustration © 2021 Alex Boersma.

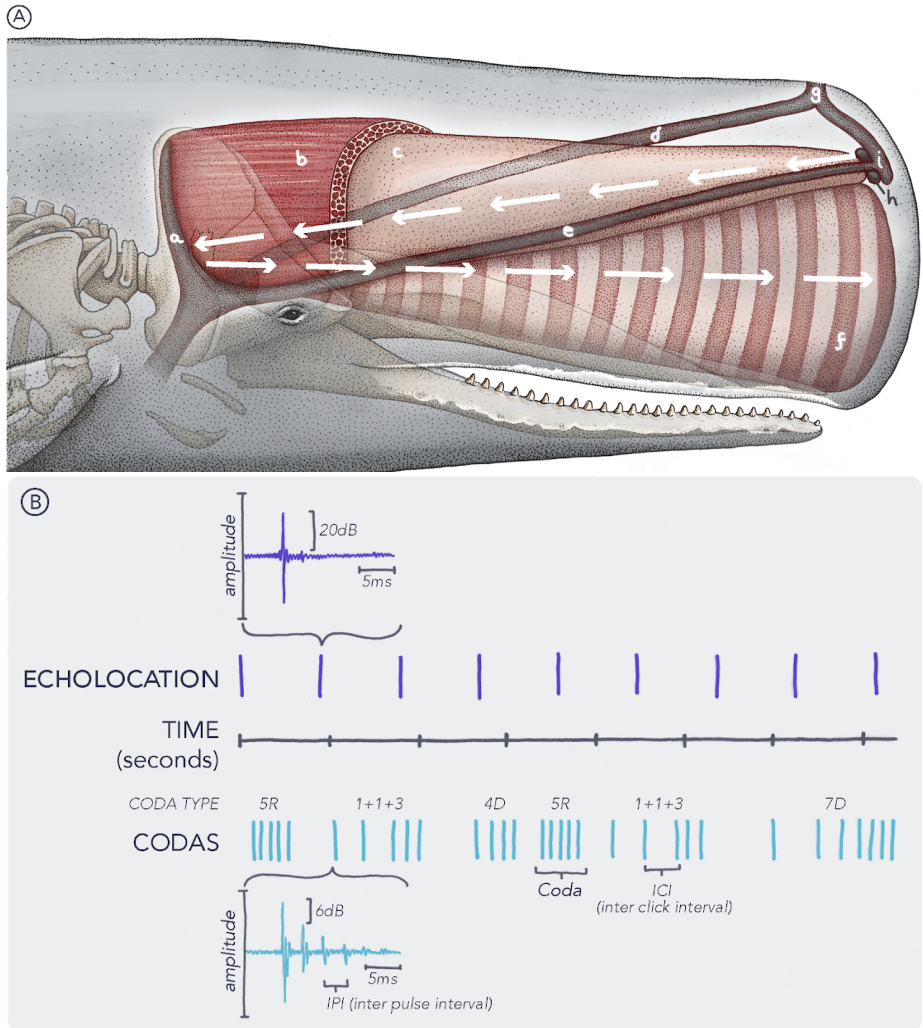

**Figure 2: Sperm whale bioacoustic system.** *A: Sperm whale head contains the spermaceti organ (**c**), a cavity filled with almost 2,000 litres of wax-like liquid, and the junk compartment (**f**), comprising a series of wafer-like bodies believed to act as acoustic lenses. The spermaceti organ and junk act as two connected tubes, forming a bent, conical horn of about 10m in length and 0.8m aperture in large mature males. The sound emitted by the phonic lips (**i**) in the front of the head is focused by traveling through the bent horn, producing a flat wavefront at the exit surface. B: Typical temporal structure of sperm whale echolocation and coda clicks. Echolocation signals are produced with consistent inter-click intervals (of approximately 0.4 sec) while coda clicks are arranged in stereotypical sequences called 'codas' lasting less than 2 sec. Codas are characterized by the different number of constituent clicks and the intervals between them (called inter-click intervals or ICIs). Codas are typically produced in multiparty exchanges that can last from about 10 seconds to over half an hour. Each click, in turn, presents itself as a sequence of equally-spaced pulses, with inter-pulse interval (IPI) of an order of 3-4 msec in an adult female, which is the result of the sound reflecting within the spermaceti organ.* Illustration © 2021 Alex Boersma.

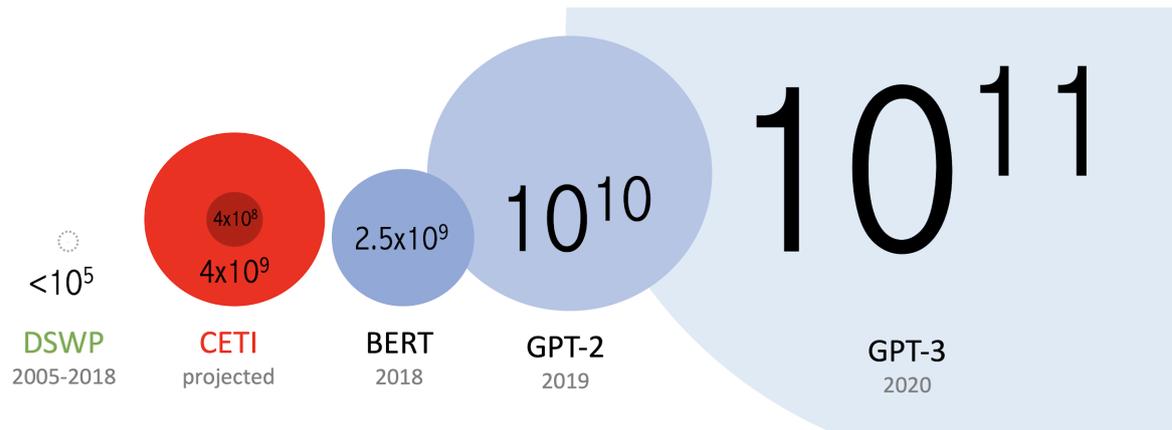

**Figure 3:** *Comparative size of datasets used for training NLP models* (represented by the circle area). GPT-3 is only partially visible, while the dataset of the Dominica Sperm Whale Project is a tiny dot on this plot (located at the center of the dashed circle). Shown in red is the estimated size of a new dataset planned to be collected in Dominica by Project CETI, an interdisciplinary initiative for cetacean communication interpretation. The estimate is based on the assumption of nearly continuous monitoring of 50-400 whales. The estimate assumes 75-80% of their vocalizations constituting echolocation clicks, and 20-25% being coda clicks. A typical Caribbean whale coda has 5 clicks and lasts 4 sec (including a silence between two subsequent codas), yielding a rate of 1.25 clicks/sec. Overall, we estimate it would be possible to collect between 400M and 4B clicks per year as a longitudinal and continuous recording of bioacoustic signals as well as detailed behavior and environmental data.

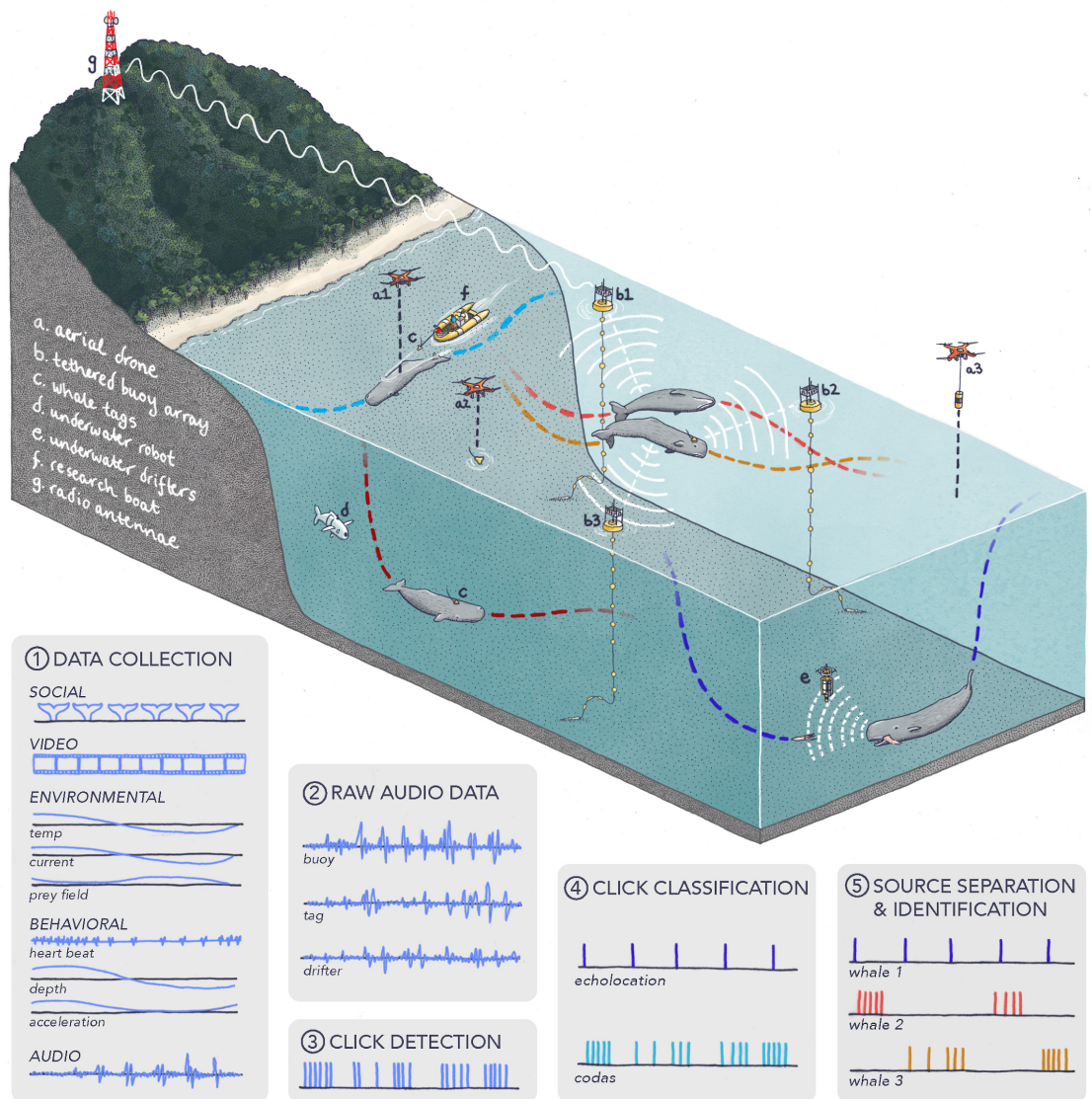

**Figure 4:** Schematic of *whale bioacoustic data collection with multiple data sources by several classes of assets. These include tethered buoy arrays (**b**), which track the whales in a large area in real-time by continuously transmitting their data to shore (**g**), floaters (**e**), and robotic fishes (**d**)Tags (**c**) attached to whales can possibly provide the most detailed bioacoustic and behavioral data. Aerial drones (**a**) can be used to assist tag deployment (**a1**), recovery (**a2**) and provide visual observation of the whales (**a3**). The collected multimodal data (**1**) has to be processed to reconstruct a social network of sperm whales. The raw acoustic data (**2**) has to be analyzed by ML algorithms to detect (**3**) and classify (**4**) clicks. Source separation and identification (**5**) algorithms would allow reconstructing multiparty conversations by attributing different clicks to the whales producing them.* Illustration © 2021 Alex Boersma.